\documentclass[a4paper]{report}
\usepackage[utf8]{inputenc}
\usepackage[T1]{fontenc}
\usepackage{RJournal}
\usepackage{amsmath,amssymb,array}
\usepackage{booktabs}

\providecommand{\tightlist}{%
  \setlength{\itemsep}{0pt}\setlength{\parskip}{0pt}}

\usepackage{longtable}

\makeatletter
 \let\@cite@ofmt\@firstofone
 \def\@biblabel#1{}
 \def\@cite#1#2{{#1\if@tempswa , #2\fi}}
\makeatother
\newlength{\cslhangindent}
\newlength{\csllabelwidth}
 {\begin{list}{}{%
  \setlength{\itemindent}{0pt}
  \setlength{\leftmargin}{0pt}
  \setlength{\parsep}{0pt}
  \ifodd #1
   \setlength{\leftmargin}{\cslhangindent}
   \setlength{\itemindent}{-1\cslhangindent}
  \fi
  \setlength{\itemsep}{#2\baselineskip}}}
 {\end{list}}
\usepackage{calc}

\usepackage{fvextra}
\fvset{breaklines=true, breakanywhere=true, fontsize=\small}

\begin{document}

%% do not edit, for illustration only
\sectionhead{Contributed research article}
\volume{18}
\volnumber{3}
\year{2026}
\month{September}

\begin{article}
  % !TeX root = RJwrapper.tex
\title{SelectionBias: An R Package for Bounding Selection Bias in Causal Estimands}

\author{by Stina Zetterstrom and Ingeborg Waernbaum}

\maketitle

\abstract{%
Selection bias may arise when there are dropouts or missing data in the analysis, or when subjects are included or excluded in the analysis based upon some selection criteria for the study population. Selection bias can jeopardize the validity of the study and a sensitivity analysis for assessing the effect of the selection is desired. Recently, there has been a surge of results for selection bias in causal inference, with several suggestions for sensitivity analyses. One method is to construct bounds for the bias. Here, we present the R package SelectionBias that can be used to calculate previously proposed bounds for selection bias for the causal risk ratio and causal risk difference for both the total and the selected populations. The first bound, derived by Smith and VanderWeele (SV), is based on values of sensitivity parameters that describe parts of the joint distribution of the outcome, treatment, selection indicator and unobserved variables. The second bound is an improved sharp bound that uses the same sensitivity parameters as the SV bound. The third bound is based solely on the observed data, and is therefore referred to as the assumption-free (AF) bound. The fourth and fifth bounds, the generalized assumption-free (GAF) and counterfactual assumption-free (CAF) bounds, utilize both the data and sensitivity parameters. The R package is illustrated with a simulated dataset that emulates a study where the effect of the zika virus on microcephaly in Brazil is investigated. Lastly, its performance and features are compared to the already existing R package EValue, highlighting situations where the two packages provide distinct advantages over each other.
}

\section{Introduction}\label{introduction}

Selecting a study population from a larger source population is a common
procedure, for example in an observational study with data from a population
register. Based on the research question, the study population is commonly
constructed using inclusion/exclusion criteria. The construction of the study
population from these selection criteria might introduce bias when estimating
the causal effect of the treatment on the outcome of interest
\citep{hernan2004structural}. This systematic error is commonly referred to as
selection bias and can occur both when interest lies in the total population
and in the selected subpopulation. Selection bias can also arise if,
for example, there are dropouts or other missing values for some individuals
in the study \citep{lu2022toward}. In an empirical study, it is often of interest to
assess the magnitude of potential biases using a sensitivity analysis. One type
of sensitivity analysis is calculating bounds for the bias, see e.g.~
\citet{huang2015bounding}, \citet{greenland2003quantifying}, \citet{flanders2019limits},
\citet{smith2019bounding}, \citet{zetterstrom2022selection}, \citet{zetterstrom2024},
\citet{zetterstrom2025investigations}, and \citet{duarte2024automated} for selection bias.

This paper presents the R package \CRANpkg{SelectionBias} which can be used to
calculate different bounds for selection bias. The ability to calculate
different bounds is attractive as these bounds are in many cases based on
assumptions that may or may not be fulfilled. One of the bounds included in
\CRANpkg{SelectionBias} is the bound presented in \citet{smith2019bounding}, hereafter
referred to as the SV bound. This bound is already included in the R package
\CRANpkg{EValue} and can also be calculated using an online calculator
\citep{smith2019bounding, smith2021multiple}. For these software tools, the user inputs
assumed values of sensitivity parameters, which are summary measures of a joint
distribution. Thus, these sensitivity parameters can be difficult to specify,
especially in the presence of many selection variables. In \CRANpkg{SelectionBias},
we also allow the user to input the model, which might be easier to specify than
the sensitivity parameters. \CRANpkg{SelectionBias} also allows the
user to assess whether the SV bound is sharp. The sharpness of a bound is a
desirable property, and recently proposed sharp bounds based on the same
sensitivity parameters as the SV bound \citep{zetterstrom2025investigations} are also
included in \CRANpkg{SelectionBias}. These bounds are referred to as the sharp
bounds. \CRANpkg{SelectionBias} also includes functions for
calculating the AF, GAF, and CAF bounds \citep{zetterstrom2022selection, zetterstrom2024}.
The sharp, AF, GAF, and CAF bounds cannot be calculated using other packages
available online. Notably, all the bounds in the
\CRANpkg{SelectionBias} package are calculated within strata of the
pre-treatment covariates. The contents of the \CRANpkg{SelectionBias} package is:

\begin{itemize}
\tightlist
\item
  \texttt{zika\_learner}: a simulated dataset inspired by the study in
  \citet{de2018association} and the zika example in \citet{smith2019bounding}. The dataset
  includes seven variables: zika virus, microcephaly, the selection variables
  birth and public hospital, the selection indicator variable and the two
  ``unmeasured'' variables living area and socioeconomic status.
\item
  \texttt{sensitivityparametersM()}: a function that calculates the sensitivity
  parameters for the SV, sharp, and GAF bounds for a model following the
  generalized M-structure (Figure \ref{fig:genM}) defined by the user.
\item
  \texttt{SVbound()}: a function that calculates the SV lower and upper bound for the
  risk ratio or risk difference in either the total or subpopulation for
  sensitivity parameters given by the user, or calculated from
  \texttt{sensitivityparametersM()}.
\item
  \texttt{sharpbound()}: a function that calculates the sharp lower and upper bound
  for the risk ratio or risk difference in either the total or subpopulation for
  sensitivity parameters given by the user, or calculated from
  \texttt{sensitivityparametersM()}.
\item
  \texttt{AFbound()}: a function that calculates the AF lower and upper bounds for
  the risk ratio or risk difference in either the total or subpopulation. The
  input can either be a dataset or probabilities calculated from the data.
\item
  \texttt{GAFbound()}: a function that calculates the GAF lower and upper bounds for
  the risk ratio or risk difference in either the total or subpopulation for
  a dataset or probabilities from the data, together with sensitivity parameters
  either given by the user, or calculated from \texttt{sensitivityparametersM()}.
\item
  \texttt{CAFbound()}: a function that calculates the CAF lower and upper bounds for
  the risk ratio or risk difference in either the total or subpopulation for
  a dataset or probabilities from the data, together with sensitivity parameters
  given by the user.
\item
  \texttt{checksharpSVbound()}: a function that evaluates if the SV bound is sharp.
\end{itemize}

The paper is outlined as follows. In Section 2, we present the causal framework,
estimands and corresponding selection biases together with an introduction of
the bounds. In Section 3, the simulated example dataset \texttt{zika\_learner} is
described. In Section 4, the R package \CRANpkg{SelectionBias} is demonstrated
using the \texttt{zika\_learner} data. The performance and features of
\CRANpkg{SelectionBias} are compared to other R packages for sensitivity
analyses, especially \CRANpkg{EValue}, in Section 5. Finally, the paper is
summarized in Section 6.

\section{Causal framework and selection bias}\label{causal-framework-and-selection-bias}

In this section, we provide a theoretical background for the bounds calculated
in the R package \CRANpkg{SelectionBias}. First, we introduce the notation and
causal framework, and second, we describe the SV, sharp, AF, GAF, and CAF bounds.

\begin{figure}

{\centering \includegraphics[width=0.35\linewidth,height=0.15\textheight]{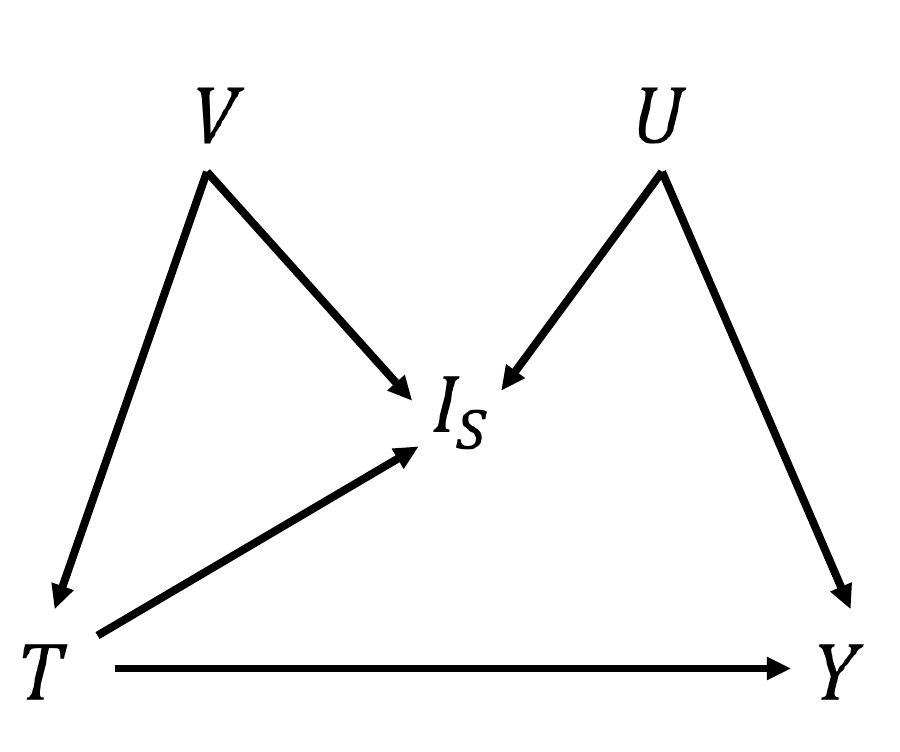} 

}

\caption{The generalized M-structure relating treatment, outcome, and selection to two unmeasured variables. Nodes represent the treatment ($T$), outcome ($Y$), selection indicator ($I_S$), and two unmeasured pre-treatment variables ($V$ and $U$); arrows represent possible direct effects between the connected variables, with $V$ pointing to both $T$ and $I_S$, and $U$ pointing to both $I_S$ and $Y$. This structure satisfies the conditional independence assumptions required for the SV, sharp, and GAF bounds, since $Y$ does not have a direct path to $I_S$ that bypasses the other variables.}\label{fig:genM}
\end{figure}

We assume \emph{K} binary selection variables, \(S_1,\dots,S_k,\dots,S_K\), where \(S_k\)
indicates if the subject passes the \emph{k}th selection criterion, from which we
define a selection indicator variable \(I_S\) such that
\begin{equation}
    I_S =
\left\{
    \begin{array}{ll}
        1  & \mbox{if } \prod\limits_{k=1}^KS_k=1  \\
        0 & \mbox{otherwise}. 
    \end{array}
\right.
    \label{eq:selInd}
\end{equation}
For a subject to be included in the study, the corresponding \(I_S\) must be equal
to 1. We consider an i.i.d. sample of size \(i=1,\ldots, n\) units from the
subpopulation such that \(I_S=1\), but henceforth suppress the index \(i\) representing
units in the sample. We also assume a binary treatment, \(T=1\) if the
unit is treated and \(T=0\) if the unit is not treated, and two corresponding
binary potential outcomes, \(Y(1)\) and \(Y(0)\) as well as an observed outcome such
that \(Y=TY(1)+(1-T)Y(0)\), i.e.~consistency is assumed \citep{DR74}. Additionally, we
assume a vector of observed pre-treatment covariates, denoted by \(X\), such that
conditional exchangeability holds in the total population: \(Y(t)\perp\hskip -6pt \perp T \mid X\),
\(t=0,1\). However, similar as to the previous literature, we suppress \(X\)
throughout to make the notation simpler and assume that all calculations are
performed within strata of the pre-treatment covariates. Lastly, we define a
vector of unobserved pre-treatment covariates, \(U\), which is part of the
assumptions in the sensitivity analysis with the SV, sharp, and GAF bounds. It is
illustrated in the generalized M-structure (Figure \ref{fig:genM}), where \(U\)
is a predictor of the outcome. In contrast, the unobserved covariates
\(U\) are not needed in the AF and CAF bounds, which impose fewer assumptions since
dependencies with \(U\) are not needed.

The causal estimands of interest are the causal risk ratio,
\(\beta_R=P(Y(1)=1)/P(Y(0)=1)\), and causal risk difference,
\(\beta_D=P(Y(1)=1)-P(Y(0)=1)\), in the total population, and the casual risk ratio,
\(\beta_{R_S}=P(Y(1)=1|I_S=1)/P(Y(0)=1|I_S=1)\), and the causal risk difference,
\(\beta_{D_S}=P(Y(1)=1|I_S=1)-P(Y(0)=1|I_S=1)\), in the subpopulation.
We also define the observed risk ratio and risk difference under
selection, \(I_S=1\), as \(\beta_R^{obs}=P(Y=1|T=1,I_S=1)/P(Y=1|T=0,I_S=1)\) and
\(\beta_D^{obs}=P(Y=1|T=1,I_S=1)-P(Y=1|T=0,I_S=1)\). These are estimated from
data, but we assume that there is no sampling variability and that \(\beta_R^{obs}\)
and \(\beta_D^{obs}\) are population quantities. The selection bias for the
risk ratio estimands, \(\beta_R\) and \(\beta_{R_S}\), are defined as the ratio
between the risk ratio and the causal estimand,
\begin{equation}
    Bias(\beta_R) = \beta_R^{obs} \big/ \beta_R,
    \label{eq:biasRR1}
\end{equation}
\begin{equation}
    Bias(\beta_{R_S}) = \beta_R^{obs} \big/ \beta_{R_S},
    \label{eq:biasRR2}
\end{equation}
and the selection bias for the risk difference estimands, \(\beta_D\) and
\(\beta_{D_S}\), are defined as the difference between the observed risk difference
and the causal estimand,
\begin{equation}
    Bias(\beta_D) = \beta_D^{obs}-\beta_D,
    \label{eq:biasRD1}
\end{equation}
\begin{equation}
    Bias(\beta_{D_S}) = \beta_D^{obs}-\beta_{D_S}.
    \label{eq:biasRD2}
\end{equation}

When the observed risk ratio or risk difference do not coincide with the causal
estimands, i.e.~when selection bias is present, bounds for the causal estimands
can be calculated and compared to the observed quantity in order to assess the
magnitude of the selection bias. We present a package that can be used to
perform sensitivity analysis for the selection bias, under various assumptions.
Bounds from the following works are considered; \citet{smith2019bounding},
\citet{zetterstrom2022selection}, \citet{zetterstrom2024}, and \citet{zetterstrom2025investigations}.
The ideas behind the bounds are shortly described here with a summary of their
properties in Table \ref{tab:bounds}. For completeness, all bounds are
presented in the Appendix.

The SV bounds rely on two versions of conditional independence assumptions
involving the outcome, selection indicator, treatment, and a vector of
unmeasured variables (see Appendix). The assumptions differ depending on the
estimand of interest. If total population estimands \(\beta_{R}\) and \(\beta_{D}\)
are of interest, conditioning on \(U\) and \(T\) must be sufficient for conditional
independence of the observed outcome \(Y\) and the selection indicator, \(I_S\). For
the case when subpopulation estimands \(\beta_{R_S}\) and \(\beta_{D_S}\) are
considered, an assumption requiring exchangeability conditional on \(U\) and \(I_S\)
is instead made. An example where both assumptions are fulfilled is the
M-structure, Figure \ref{fig:genM}, although other structures are possible.
The purpose of the different assumptions is to describe a setting where unbiased
estimates could be obtained if the unmeasured variables were observed. The SV
bounds are based on sensitivity parameters that are constructed as risk ratios
formed by the joint distribution of the outcome, treatment, selection indicator
variable and unmeasured variables, \((Y,T,I_S,U)\). The sensitivity parameters are
\[
\mathrm{RR}_{SU|ts} = \max_u \frac{p(U=u \mid I_S=s, T=t)}{p(U=u \mid I_S=1-s, T=t)},
\]
\[
\mathrm{RR}_{UY|t} = \frac{\max_u p(Y=1 \mid U=u, T=t)}{\min_u p(Y=1 \mid U=u, T=t)},
\]
and
\[
\mathrm{BF}_{ts} = \frac{\mathrm{RR}_{SU|ts} \times \mathrm{RR}_{UY|t}}{\mathrm{RR}_{SU|ts} + \mathrm{RR}_{UY|t} - 1},
\]
from which bounds for the causal risk ratio and causal risk difference can be
calculated. For the subpopulation, SV defined the sensitivity parameters

\[
\mathrm{RR}_{TU|t} = \max_u \frac{p(U=u \mid I_S=1, T=t)}{p(U=u \mid I_S=1, T=1-t)},
\]

\[
\mathrm{RR}_{UY|S=1} = \max_t \frac{\max_u p(Y=1 \mid U=u, T=t, I_S=1)}{\min_u p(Y=1 \mid U=u, T=t, I_S=1)},
\]

and

\[
\mathrm{BF}_{t} = \frac{\mathrm{RR}_{TU|t} \times \mathrm{RR}_{UY|S=1}}{\mathrm{RR}_{TU|t} + \mathrm{RR}_{UY|S=1} - 1}.
\]
More details of the sensitivity parameters and bounds can be found in the
original work. It may be difficult to interpret and specify the sensitivity
parameters, especially in the presence of multiple selection variables and
multiple unmeasured variables. Instead, it may be easier to break down the bound
into smaller pieces, and calculate the sensitivity parameters for a specified
data generating process (DGP). We facilitate such calculations in the R package
\CRANpkg{SelectionBias}.

A bound is sharp if the bias can equal the value of the bound, given the
necessary assumptions, an observed distribution, and correctly specified
sensitivity parameters. Since different bounds require different assumptions
and information, two bounds for the same causal estimand can be simultaneously
sharp if each is valid under its own requirements. The SV bounds are only sharp
under certain conditions, i.e.~sometimes the SV bounds are too conservative and
are larger than the bias can actually be. As an alternative, the sensitivity
parameters have been used to construct improved bounds that are generally sharp
\citep{zetterstrom2025investigations}, referred to as the sharp bounds. However, it
should be noted that they do in some cases require some additional knowledge of
the selection probability.

The AF bounds are constructed solely from the data and have the advantage that
no additional assumptions regarding the causal model are needed, only that
\(P(I_S=1)\) is known. The bounds are constructed as the minimum and maximum
possible values of the causal estimand, \(\beta^{min}\) and \(\beta^{max}\). These
values are then used as lower and upper bounds for the causal
estimand. As the bounds also demonstrate the maximum selection bias,
they can be used to assess whether other bounds are informative: if another
bound is greater than the AF bound, it is not informative. However, if the
treatment or outcome is rare, the AF bounds for the risk ratios can be very
wide, and practically non-informative. When this is the case, and additional
knowledge is available, it is instead advisable to use other bounds that
take that knowledge into account.

The GAF bounds are similar to the AF bounds in so far they also utilize the
observed data, but they also rely on the same conditional independence
assumptions as the SV bound (see Appendix), which results in tighter bounds
compared to the AF bounds. The sensitivity parameters are probabilities instead
of risk ratios, which can in some instances be easier to specify.
The sensitivity parameters are defined as
\[
    m_T=\min_{t,u}P(Y=1|T=t,U=u)
\]
and
\[
    M_T= \max_{t,u}P(Y=1|T=t,U=u)
\]
for the total population, and

\[
    m_S=\min_{t,u}P(Y=1|T=t,U=u,S=1)
\]
and
\[
    M_S =\max_{t,u}P(Y=1|T=t,U=u,S=1)
\]
for the subpopulation.

The CAF bounds are an alternative when the conditional independence assumptions
are not fulfilled, but they require the user to specify counterfactual
probabilities, \(P(Y(t)=1|T=1-t,I_S=1)\). The sensitivity parameters are defined as
\[
    m'_T=\min_t P(Y=1|T=t,I_S=0)
\]
and
\[
    M'_T=\max_t P(Y=1|T=t,I_S=0),
\]
for the total population, and

\[
    m_S'=\min_t P(Y(t)=1|T=1-t,I_S=1)
\]
and
\[
    M_S'=\max_t P(Y(t)=1|T=1-t,I_S=1),
\]
for the subpopulation. The AF bounds can of course be used if the conditional
independence assumption is not fulfilled, but the CAF bounds are less conservative.
The GAF and CAF bounds are equal to the AF bound when the most extreme values on
sensitivity parameters are used.

\begin{longtable}[]{@{}
  >{\centering\arraybackslash}p{(\linewidth - 8\tabcolsep) * \real{0.2647}}
  >{\centering\arraybackslash}p{(\linewidth - 8\tabcolsep) * \real{0.2206}}
  >{\centering\arraybackslash}p{(\linewidth - 8\tabcolsep) * \real{0.0882}}
  >{\centering\arraybackslash}p{(\linewidth - 8\tabcolsep) * \real{0.1618}}
  >{\centering\arraybackslash}p{(\linewidth - 8\tabcolsep) * \real{0.2647}}@{}}
\caption{\label{tab:bounds} Summary of the different bounds.}\tabularnewline
\toprule\noalign{}
\begin{minipage}[b]{\linewidth}\centering
Bound
\end{minipage} & \begin{minipage}[b]{\linewidth}\centering
Unknown sensitivity parameters
\end{minipage} & \begin{minipage}[b]{\linewidth}\centering
Uses data
\end{minipage} & \begin{minipage}[b]{\linewidth}\centering
Additional assumptions
\end{minipage} & \begin{minipage}[b]{\linewidth}\centering
Specify counterfactual probabilities
\end{minipage} \\
\midrule\noalign{}
\endfirsthead
\toprule\noalign{}
\begin{minipage}[b]{\linewidth}\centering
Bound
\end{minipage} & \begin{minipage}[b]{\linewidth}\centering
Unknown sensitivity parameters
\end{minipage} & \begin{minipage}[b]{\linewidth}\centering
Uses data
\end{minipage} & \begin{minipage}[b]{\linewidth}\centering
Additional assumptions
\end{minipage} & \begin{minipage}[b]{\linewidth}\centering
Specify counterfactual probabilities
\end{minipage} \\
\midrule\noalign{}
\endhead
\bottomrule\noalign{}
\endlastfoot
Smith and VanderWeele (SV) & \(\checkmark\) & & \(\checkmark\) & \\
Sharp & \(\checkmark\) & \(\checkmark\) & \(\checkmark\) & \\
Assumption-free (AF) & & \(\checkmark\) & & \\
Generalized assumption-free (GAF) & \(\checkmark\) & \(\checkmark\) & \(\checkmark\) & \\
Counterfactual assumption-free (CAF) & \(\checkmark\) & \(\checkmark\) & & \(\checkmark\) \\
\end{longtable}

\section{The simulated data set}\label{the-simulated-data-set}

For the purpose of illustration of the bounds, we construct a simulated dataset,
\texttt{zika\_learner}. This is inspired by a case-control study that investigates the
effect of zika virus on microcephaly \citep{de2018association} and the numerical zika
example introduced in \citet{smith2019bounding}. The data can be loaded by

\begin{verbatim}
data("zika_learner", package = "SelectionBias")
\end{verbatim}

\begin{figure}

{\centering \includegraphics[width=0.35\linewidth]{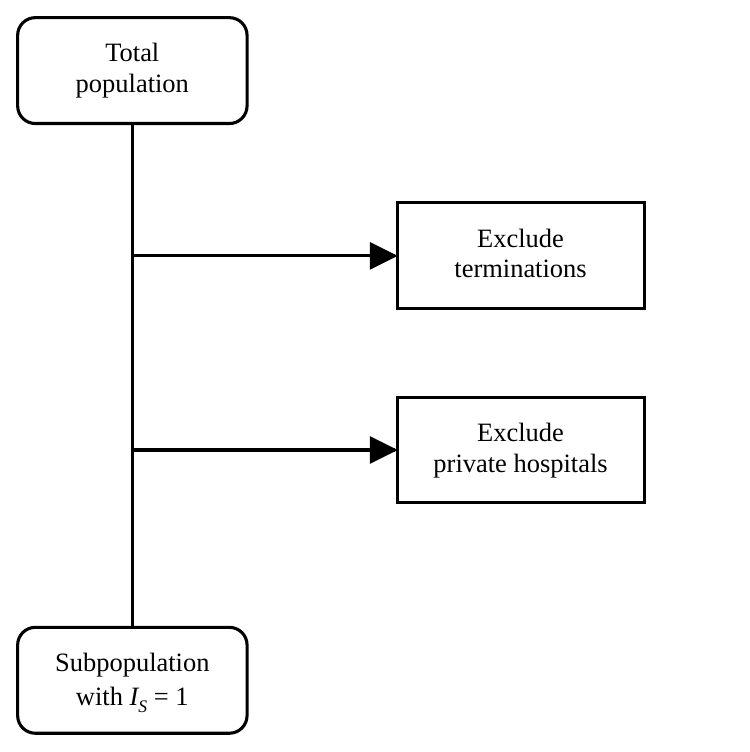} \includegraphics[width=0.35\linewidth]{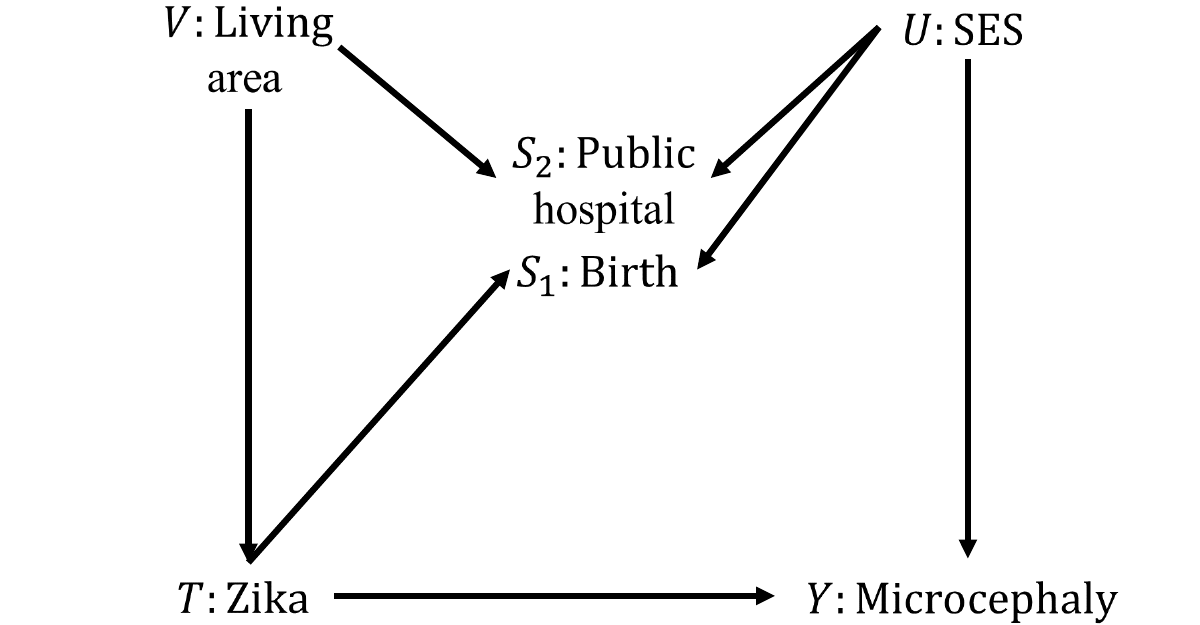} 

}

\caption{Selection process and causal structure for the simulated dataset. Left: flow chart of the two-step selection from the total population. Pregnancies ending in termination are excluded first, followed by births occurring at private hospitals, leaving the subpopulation with $I_S=1$. Right: the corresponding causal DAG, where living area ($V$) and socioeconomic status ($U$) are unmeasured variables affecting the two selection steps, birth ($S_1$) and public hospital ($S_2$), while zika infection ($T$) affects microcephaly ($Y$) directly. The DAG shows that both selection steps depend on the same unmeasured variables that also relate to the outcome, which is why selection bias can arise even though $T$ and $Y$ are not confounded in the total population.}\label{fig:flow}
\end{figure}

The zika example in \citet{smith2019bounding} covers the case of a single selection,
and we expand the example with a second selection variable to demonstrate
multiple inclusion criteria which are common in empirical applications. The two
selections are the binary variables \texttt{birth} and \texttt{public\ hospital}, and the
selection process is described in Figure \ref{fig:flow} (left). In the original
study, pre-treatment covariates for both the mother and the infant were included
in order to control for confounding \citep{de2018association}, however, we only
consider one stratum of the covariates and thus exclude all observed pre-treatment
covariates in the simulated dataset. The causal model of the dataset is given in
Figure \ref{fig:flow} (right). All variables are binary and generated from the
binomial distribution such that the prevalences of the variables, and strengths
of dependencies between them, mimic the real empirical data and the assumed
values for the sensitivity parameters in \citet{smith2019bounding}. The causal
dependencies are generated by the logistic models described in Table
\ref{tab:dgp}. Even though the dataset is inspired by a case-control study, the
simulated dataset emulates a register with 5000 observations, similar to an
observational study of zika virus \citep{lebov2019international}. The variables
included are:

\begin{itemize}
\tightlist
\item
  \texttt{urban} (\(V\)). A binary, unobserved variable indicating whether the subject
  lives in an urban area (\(V=1\)) or not (\(V=0\)). The probability of
  living in an urban area is set to 0.85 following numbers from the World Bank \citep{urban}.
\item
  \texttt{ses} (\(U\)). A binary variable indicating
  socioeconomic status (high vs low), with probability set to 0.5.
\item
  \texttt{zika} (\(T\)). In 2016 there were approximately 34000 possible cases of zika virus
  during pregnancy \citep{de2017infection}, and there are approximately 2.9 million
  births per year in Brazil \citep{diniz2019understanding}. The risk of getting zika was
  higher in urban areas \citep{ali2017environmental}. Thus, the prevalence of zika
  in the simulated dataset is around 1\% with a positive impact of \(V\), see
  Table \ref{tab:dgp} for details.
\item
  \texttt{mic\_ceph} (\(Y\)). The estimated prevalence of microcephaly was 74
  cases per 10000 births and the controlled odds ratio of zika virus on
  microcephaly was estimated at 73.1 \citep{de2018association}. We mimic these
  values with an overall prevalence of 0.8\% and a risk ratio of 74.5 among
  the subpopulation after two selections, see Table \ref{tab:dgp} for details.
\item
  \texttt{birth} (\(S_1\)). Pregnancies that ended in a live or still birth were
  included in the study, and pregnancies that ended in a miscarriage or an
  abortion were excluded. The probability of a pregnancy ending in birth is
  assumed to be affected both by zika virus infection and socioeconomic
  status. The number of births per year in Brazil is approximately 2.9 million
  and the number of unregulated abortions per year is estimated at 500000
  \citep{malta2019abortion}. The prevalence of birth in the simulated dataset is
  0.86, with a strong negative impact of zika virus, and a positive impact of
  socioeconomic status, see Table \ref{tab:dgp} for details.
\item
  \texttt{hospital} (\(S_2\)). Births that occurred in public hospitals were
  included in the study, and births that occurred in private hospitals were
  excluded. A majority of the population in Brazil visits public hospitals,
  and here it is assumed that it is strongly affected by socioeconomic status
  and weakly affected by living area, see Table \ref{tab:dgp} for details.
\end{itemize}

\begin{longtable}[]{@{}
  >{\centering\arraybackslash}p{(\linewidth - 4\tabcolsep) * \real{0.4149}}
  >{\centering\arraybackslash}p{(\linewidth - 4\tabcolsep) * \real{0.3723}}
  >{\raggedleft\arraybackslash}p{(\linewidth - 4\tabcolsep) * \real{0.2128}}@{}}
\caption{\label{tab:dgp} Data generating process for the dataset \texttt{zika\_learner}. Models
generating causal dependencies are logistic, \(g(X'\theta)\), for predictor
variables \(X\) (including a constant 1 for the intercept) and model parameter
\(\theta\) (including the intercept).}\tabularnewline
\toprule\noalign{}
\begin{minipage}[b]{\linewidth}\centering
Model
\end{minipage} & \begin{minipage}[b]{\linewidth}\centering
Coefficients (\(\theta\))/Proportions
\end{minipage} & \begin{minipage}[b]{\linewidth}\raggedleft
Function argument
\end{minipage} \\
\midrule\noalign{}
\endfirsthead
\toprule\noalign{}
\begin{minipage}[b]{\linewidth}\centering
Model
\end{minipage} & \begin{minipage}[b]{\linewidth}\centering
Coefficients (\(\theta\))/Proportions
\end{minipage} & \begin{minipage}[b]{\linewidth}\raggedleft
Function argument
\end{minipage} \\
\midrule\noalign{}
\endhead
\bottomrule\noalign{}
\endlastfoot
\(P(V=1)\) & \(0.85\) & \texttt{Vval} \\
\(P(U=1)\) & \(0.50\) & \texttt{Uval} \\
\(P(T=1|V)=g(V'\theta_T)\) & \((-6.20,1.75)\) & \texttt{Tcoef} \\
\(P(Y=1|T,U)=g[(T,U)'\theta_{Y}]\) & \((-5.20,5.00,-1.00)\) & \texttt{Ycoef} \\
\(P(S_1=1|V,U,T)=g[(V,U,T)'\theta_{S1}]\) & \((1.20,0.00,2.00,-4.00)\) & \texttt{Scoef} \\
\(P(S_2=1|V,U,T)=g[(V,U,T)'\theta_{S2}]\) & \((2.20,0.50,-2.75,0.00)\) & \texttt{Scoef} \\
\end{longtable}

The unobserved variables, \texttt{urban} (\(V\)) and \texttt{ses} (\(U\)) are constructed such
that the true sensitivity parameters in our DGP approximately match the assumed
values in \citet{smith2019bounding}. In the simulated dataset, the proportions of the
variables presented in Tables \ref{tab:total}-\ref{tab:sel2} reflect this. The proportions are calculated both by treatment status
(zika virus infection) and overall, for the total dataset, the
subset with \(S_1=1\) and the subset with \(S_1=1\) and \(S_2=1\).

\begin{longtable}[]{@{}cccc@{}}
\caption{\label{tab:total} Proportions and number of subjects for the simulated dataset, by treatment (zika) status and overall.}\tabularnewline
\toprule\noalign{}
& Not zika infected & Zika infected & Overall \\
\midrule\noalign{}
\endfirsthead
\toprule\noalign{}
& Not zika infected & Zika infected & Overall \\
\midrule\noalign{}
\endhead
\bottomrule\noalign{}
\endlastfoot
Number of subjects & 4939 & 61 & 5000 \\
Microcephaly & 0.003 & 0.361 & 0.008 \\
Living area & 0.849 & 0.951 & 0.850 \\
SES & 0.499 & 0.426 & 0.498 \\
\end{longtable}

\begin{longtable}[]{@{}cccc@{}}
\caption{\label{tab:sel1} Proportions and number of subjects for the simulated dataset, by treatment (zika) status and overall, after the first selection.}\tabularnewline
\toprule\noalign{}
& Not zika infected & Zika infected & Overall \\
\midrule\noalign{}
\endfirsthead
\toprule\noalign{}
& Not zika infected & Zika infected & Overall \\
\midrule\noalign{}
\endhead
\bottomrule\noalign{}
\endlastfoot
Number of subjects & 4268 & 11 & 4279 \\
Microcephaly & 0.003 & 0.273 & 0.004 \\
Living area & 0.845 & 1.000 & 0.846 \\
SES & 0.556 & 0.818 & 0.557 \\
\end{longtable}

For the DGP, the causal estimand is \(\beta_R=90.7\) in the total population,
\(\beta_{R_{S}}=92.3\) in the subset of \(S_1=1\), and \(\beta_{R_{S}}=88.1\) in the
subset of \(S_1=1\) and \(S_2=1\). The observed risk ratios for the dataset are
\(\beta_R^{obs}=89.5\) (after one selection), and \(\beta_R^{obs}=74.5\) (after two
selections). The dataset and DGP can be used to apply the functions in \CRANpkg{SelectionBias}.

\begin{longtable}[]{@{}cccc@{}}
\caption{\label{tab:sel2} Proportions and number of subjects for the simulated dataset, by treatment (zika) status and overall, after both selections.}\tabularnewline
\toprule\noalign{}
& Not zika infected & Zika infected & Overall \\
\midrule\noalign{}
\endfirsthead
\toprule\noalign{}
& Not zika infected & Zika infected & Overall \\
\midrule\noalign{}
\endhead
\bottomrule\noalign{}
\endlastfoot
Number of subjects & 2869 & 7 & 2876 \\
Microcephaly & 0.004 & 0.286 & 0.005 \\
Living area & 0.858 & 1.000 & 0.858 \\
SES & 0.382 & 0.714 & 0.382 \\
\end{longtable}

\section{R package SelectionBias}\label{r-package-selectionbias}

In this section, we describe the package \CRANpkg{SelectionBias}. The different
functions are illustrated using either the simulated dataset \texttt{zika\_learner}, or
the model structure that the dataset is simulated from (see Table
\ref{tab:dgp}). The package \CRANpkg{SelectionBias} is available under a
GPL-compatible license from the Comprehensive R Archive Network (CRAN) at
\url{https://cran.r-project.org/package=SelectionBias}, and can be installed and
loaded using:

\begin{verbatim}
install.packages("SelectionBias")
\end{verbatim}

\begin{verbatim}
library(SelectionBias)
\end{verbatim}

\subsection{\texorpdfstring{\texttt{sensitivityparametersM()}}{sensitivityparametersM()}}\label{sensitivityparametersm}

The sensitivity parameters for the SV, sharp, and GAF bounds are calculated for
the generalized M-structure illustrated in Figure \ref{fig:genM} in the function
\texttt{sensitivityparametersM()}. The input in the function is the model
structure, and the observed probabilities of the outcome, \(P(Y=1|T=t,I_S=1)\),
\(t=0,1\). The code and the output are:

\begin{verbatim}
# SV bound
sensparSV = sensitivityparametersM(whichEst = "RR_tot",
                   whichBound = "SV",
                   Vval = matrix(c(1, 0, 0.85, 0.15), ncol = 2),
                   Uval = matrix(c(1, 0, 0.5, 0.5), ncol = 2),
                   Tcoef = c(-6.2, 1.75),
                   Ycoef = c(-5.2, 5.0, -1.0),
                   Scoef = matrix(c(1.2, 2.2, 0.0, 0.5,
                                    2.0, -2.75, -4.0, 0.0),
                                  ncol = 4),
                   Mmodel = "L",
                   pY1_T1_S1 = 0.286,
                   pY1_T0_S1 = 0.004)
print(sensparSV)
\end{verbatim}

\begin{verbatim}
#> $parameter
#>  [1] "BF_11"     "BF_00"     "BF_10"     "BF_01"     "RR_UY|T=1" "RR_UY|T=0"
#>  [7] "RR_SU|11"  "RR_SU|00"  "RR_SU|10"  "RR_SU|01" 
#> 
#> $value
#>  [1] 1.2102 1.3532 1.3208 1.3895 1.9448 2.7089 1.5566 1.7058 1.9998 1.7998
#> 
#> attr(,"class")
#> [1] "sensparams"
#> attr(,"row.names")
#>  [1]  1  2  3  4  5  6  7  8  9 10
\end{verbatim}

\begin{verbatim}
# Sharp bound
sensparSharp = sensitivityparametersM(whichEst = "RR_tot",
                   whichBound = "sharp",
                   Vval = matrix(c(1, 0, 0.85, 0.15), ncol = 2),
                   Uval = matrix(c(1, 0, 0.5, 0.5), ncol = 2),
                   Tcoef = c(-6.2, 1.75),
                   Ycoef = c(-5.2, 5.0, -1.0),
                   Scoef = matrix(c(1.2, 2.2, 0.0, 0.5,
                                    2.0, -2.75, -4.0, 0.0),
                                  ncol = 4),
                   Mmodel = "L",
                   pY1_T1_S1 = 0.286,
                   pY1_T0_S1 = 0.004)

print(sensparSharp)
\end{verbatim}

\begin{verbatim}
#> $parameter
#>  [1] "BF_11"     "BF_00"     "BF_10"     "BF_01"     "RR_UY|T=1" "RR_UY|T=0"
#>  [7] "RR_SU|11"  "RR_SU|00"  "RR_SU|10"  "RR_SU|01" 
#> 
#> $value
#>  [1] 1.2102 1.3532 1.3208 1.3895 1.9448 2.7089 1.5566 1.7058 1.9998 1.7998
#> 
#> attr(,"class")
#> [1] "sensparams"
#> attr(,"row.names")
#>  [1]  1  2  3  4  5  6  7  8  9 10
\end{verbatim}

\begin{verbatim}
# GAF bound
sensparGAF = sensitivityparametersM(whichEst = "RR_tot",
                   whichBound = "GAF",
                   Vval = matrix(c(1, 0, 0.85, 0.15), ncol = 2),
                   Uval = matrix(c(1, 0, 0.5, 0.5), ncol = 2),
                   Tcoef = c(-6.2, 1.75),
                   Ycoef = c(-5.2, 5.0, -1.0),
                   Scoef = matrix(c(1.2, 2.2, 0.0, 0.5,
                                    2.0, -2.75, -4.0, 0.0),
                                  ncol = 4),
                   Mmodel = "L",
                   pY1_T1_S1 = 0.286,
                   pY1_T0_S1 = 0.004)

print(sensparGAF)
\end{verbatim}

\begin{verbatim}
#> $parameter
#> [1] "m_T" "M_T"
#> 
#> $value
#> [1] 0.0020 0.4502
#> 
#> attr(,"class")
#> [1] "sensparams"
#> attr(,"row.names")
#> [1] 1 2
\end{verbatim}

The first argument is \texttt{whichEst}, where the user inputs the causal estimand of
interest. It must be one of the four \texttt{"RR\_tot"}, \texttt{"RD\_tot"}, \texttt{"RR\_sub"} or
\texttt{"RD\_sub"}. The second argument is \texttt{whichBound}, where the user inputs the bound
they want to use, either \texttt{"SV"}, \texttt{"sharp"}, or \texttt{"GAF"}. Third, the argument \texttt{Vval}
takes the matrix for \emph{V} as input. The first column contains the values that \emph{V}
can take, and the second column contains the corresponding probabilities. In
this example, \emph{V} is binary, so the first two elements in the matrix are 1 and 0.
However, any discrete \emph{V} can be used. An approximation of a continuous \emph{V} can
be used, if it is discretized. The fourth argument is \texttt{Uval}, which takes the
matrix for \emph{U} as input. The matrix \emph{U} has a similar structure as \emph{V}. The
fifth argument is \texttt{Tcoef}, containing the coefficients used in the model for \emph{T}.
The first entry in \texttt{Tcoef} is the intercept of the model, and the second the
slope for \emph{V}. The sixth argument is \texttt{Ycoef}, containing the coefficient vector
for the outcome model, where the first entry is the intercept, the second the
slope coefficient for \emph{T} and third is the slope coefficient for \emph{U}. The seventh
argument is \texttt{Scoef}, the coefficient matrix for the selection variables. The
number of rows is equal to the number of selection variables, and the number of
columns is equal to four. The columns represent the intercept, and slope
coefficients for \emph{V}, \emph{U} and \emph{T}, respectively. A summary of the code
notation is seen in the last column of Table \ref{tab:dgp}. The eighth argument
is \texttt{Mmodel}, which indicates whether the models in the M-structure are probit
(\texttt{Mmodel\ =\ "P"}) or logit (\texttt{Mmodel\ =\ "L"}). The ninth and tenth arguments are
\texttt{pY1\_T1\_S1} and \texttt{pY1\_T0\_S1}. They are the observed probabilities
\(P(Y=1|T=1,I_S=1)\) and \(P(Y=1|T=0,I_S=1)\). The output is the sensitivity
parameters for the chosen bound.

In the zika example, the estimand of interest is the risk ratio in the
total population, \texttt{whichEst\ =\ "RR\_tot"}, the DGP is found in Table \ref{tab:dgp},
logistic models are used in the DGP, and the probabilities are found in Table
\ref{tab:sel2}. For the SV and sharp bounds, the output is \(RR_{UY|T=1}=1.9448\),
\(RR_{UY|T=0}=2.7089\), \(RR_{SU|11}=1.5566\), \(RR_{SU|00}=1.7058\),
\(RR_{SU|10}=1.9998\), and \(RR_{SU|01}=1.7998\), which gives \(BF_{11}=1.2102\) and
\(BF_{00}=1.3532\), \(BF_{10}=1.3208\), and \(BF_{01}=1.3895\). For the GAF bound,
the output is \(M_T=0.4502\) and \(m_T=0.002\).

\subsection{\texorpdfstring{\texttt{SVbound()}}{SVbound()}}\label{svbound}

The SV bound can be calculated using the function \texttt{SVbound()}. The first
argument is \texttt{whichEst}, indicating the causal estimand of interest (\texttt{"RR\_tot"},
\texttt{"RD\_tot"}, \texttt{"RR\_sub"} or \texttt{"RD\_sub"}). The second argument is \texttt{sens}, which is
either output from \texttt{sensitivityparametersM()}, a named list, or a data frame with
column names parameter and value, which includes the sensitivity parameters.
This input is optional as the sensitivity parameters can also be specified
directly in the function with the below arguments. The third and fourth arguments
are the observed conditional probabilities \(P(Y=1|T=1,I_S=1)\) and
\(P(Y=1|T=0,I_S=1)\), which are needed to calculate bounds for the causal
estimands and not the selection bias itself. The optional fifth and sixth
arguments are the treatment probabilities \(P(T=1|I_S=1)\) and \(P(T=0|I_S=1)\).
These are only needed for the alternative SV bound for the causal risk difference
in the subpopulation \citep{zetterstrom2025investigations}. If they are not included,
the original SV bound for the risk difference is calculated. The subsequent
arguments are the sensitivity parameters provided by the user. The default value
for all sensitivity parameters is \texttt{NULL}, and the user must then specify numeric
values on the sensitivity parameters that are necessary for the bound for the
chosen estimand. These arguments are also optional in case the argument \texttt{sens}
is used. The sensitivity parameters can either be calculated using the function
\texttt{sensitivityparametersM()}, or found elsewhere. For sensitivity parameters found
elsewhere, \texttt{SVbound()} is not restricted to the generalized M-structure.
However, the necessary assumptions for the SV bound must still be fulfilled
\citep{smith2019bounding}. The output is the SV bound. The code and output are:

\begin{verbatim}
SVbound(whichEst = "RR_tot",
        pY1_T1_S1 = 0.286,
        pY1_T0_S1 = 0.004,
        RR_UY_T1 = 1.9448,
        RR_UY_T0 = 2.7089,
        RR_SU_11 = 1.5566,
        RR_SU_00 = 1.7058,
        RR_SU_10 = 1.9998,
        RR_SU_01 = 1.7998)
\end{verbatim}

\begin{verbatim}
#>      [,1]             [,2]  
#> [1,] "SV lower bound" 43.66 
#> [2,] "SV upper bound" 131.22
\end{verbatim}

As before in the zika example, the causal estimand is the risk ratio in the
total population, \texttt{whichEst\ =\ "RR\_tot"}. The sensitivity parameters are
\(RR_{UY|T=1}=1.9448\), \(RR_{UY|T=0}=2.7089\), \(RR_{SU|11}=1.5566\),
\(RR_{SU|00}=1.7058\),\(RR_{SU|10}=1.9998\), and \(RR_{SU|01}=1.7998\), calculated
above in \texttt{sensitivityparametersM()}, which gives a lower SV bound equal to 43.66
and a higher SV bound equal to 131.22. Alternatively, the output from the function
\texttt{sensitivityparametersM()} can be used directly as input which gives the code:

\begin{verbatim}
SVbound(whichEst = "RR_tot",
        sens = sensparSV,
        pY1_T1_S1 = 0.286,
        pY1_T0_S1 = 0.004)
\end{verbatim}

\begin{verbatim}
#>      [,1]             [,2]  
#> [1,] "SV lower bound" 43.66 
#> [2,] "SV upper bound" 131.22
\end{verbatim}

This approach gives the same bounds as specifying the sensitivity parameters
manually.

\subsection{\texorpdfstring{\texttt{sharpbound()}}{sharpbound()}}\label{sharpbound}

The sharp bound can be calculated using the function \texttt{sharpbound()}. The input
arguments are very similar to that of \texttt{SVbound()}. The first argument is
\texttt{whichEst}, indicating the causal estimand of interest (\texttt{"RR\_tot"},
\texttt{"RD\_tot"}, \texttt{"RR\_sub"} or \texttt{"RD\_sub"}). The second argument is \texttt{sens}, which is
either output from \texttt{sensitivityparametersM()}, a named list, or a data frame with
column names parameter and value, which includes the sensitivity parameters.
This input is optional as the sensitivity parameters can also be specified
directly in the function with the below arguments. The third and fourth arguments
are the observed conditional probabilities \(P(Y=1|T=1,I_S=1)\) and
\(P(Y=1|T=0,I_S=1)\), which are needed to calculate bounds for the causal
estimands and not the selection bias itself. The fifth and sixth
arguments are the treatment probabilities \(P(T=1|I_S=1)\) and \(P(T=0|I_S=1)\).
These are only needed for the bounds for the causal estimands in the
subpopulation. The seventh and eighth arguments are the probabilities
\(P(I_S=1|T=1)\) and \(P(I_S=1|T=0)\) which are only needed for the bounds for the
causal estimands in the total population. If they are unknown, they can be set
to 0 for more conservative bounds. The subsequent arguments are the
sensitivity parameters provided by the user. The default value for all
sensitivity parameters is \texttt{NULL}, and the user must then specify numeric values
on the sensitivity parameters that are necessary for the bound for the chosen
estimand. These arguments are also optional in case the argument \texttt{sens} is used.
The sensitivity parameters can either be calculated using the function
\texttt{sensitivityparametersM()}, or found elsewhere. For sensitivity parameters found
elsewhere, \texttt{sharpbound()} is not restricted to the generalized M-structure.
However, the necessary assumptions for the sharp bound must still be fulfilled
\citep{zetterstrom2025investigations}. The output is the sharp bound. The code and
output with the two options to specify the sensitivity parameters are:

\begin{verbatim}
sharpbound(whichEst = "RR_tot",
        pY1_T1_S1 = 0.286,
        pY1_T0_S1 = 0.004,
        pS1_T1 = 0.11,
        pS1_T0 = 0.58,
        RR_UY_T1 = 1.9448,
        RR_UY_T0 = 2.7089,
        RR_SU_11 = 1.5566,
        RR_SU_00 = 1.7058,
        RR_SU_10 = 1.9998,
        RR_SU_01 = 1.7998)
\end{verbatim}

\begin{verbatim}
#>      [,1]                [,2] 
#> [1,] "Sharp lower bound" 73.91
#> [2,] "Sharp upper bound" 78.99
\end{verbatim}

\begin{verbatim}
sharpbound(whichEst = "RR_tot",
        sens = sensparSharp,
        pY1_T1_S1 = 0.286,
        pY1_T0_S1 = 0.004,
        pS1_T1 = 0.11,
        pS1_T0 = 0.58)
\end{verbatim}

\begin{verbatim}
#>      [,1]                [,2] 
#> [1,] "Sharp lower bound" 73.91
#> [2,] "Sharp upper bound" 78.99
\end{verbatim}

As before in the zika example, the causal estimand is the risk ratio in the
total population, \texttt{whichEst\ =\ "RR\_tot"}. The observed probabilities come from
the dataset, and the sensitivity parameters are \(RR_{UY|T=1}=1.9448\),
\(RR_{UY|T=0}=2.7089\), \(RR_{SU|11}=1.5566\), \(RR_{SU|00}=1.7058\),
\(RR_{SU|10}=1.9998\), and \(RR_{SU|01}=1.7998\), calculated above in
\texttt{sensitivityparametersM()}, which gives a lower sharp bound equal to 73.91 and a
higher sharp bound equal to 78.99.

\subsection{\texorpdfstring{\texttt{checksharpSVbound()}}{checksharpSVbound()}}\label{checksharpsvbound}

The sharpness of the SV bound can be evaluated using \texttt{checksharpSVbound()}
\citep{zetterstrom2025investigations}. The first argument is
\texttt{whichEst}, indicating the causal estimand of interest (\texttt{"RR\_tot"}, \texttt{"RR\_sub"}
or \texttt{"RD\_sub"}). Note that the SV bound for the risk difference in the total
population is not sharp. The second argument is \texttt{sens}, which is either output
from \texttt{sensitivityparametersM()}, a named list or a data frame with
column names parameter and value, which includes the bounding factors (\(BF\)).
This input is optional as the bounding factors can also be specified
directly in the function with the argument below. The third argument is \texttt{BF},
the bounding factors \(BF_{00}\) and \(BF_{10}\) for the total population, and \(BF_0\)
and \(BF_1\) for the subpopulation. The fourth argument is \texttt{pY1}, the probabilities
\(P(Y=1|T=1,I_S=1)\) and \(P(Y=1|T=0,I_S=1)\). Note that the order of the bounding
factors and probabilities matters. The output is two strings stating whether the
lower and upper SV bound are sharp or not. Note that the SV bound for the risk
ratio in the total population and risk difference in the subpopulation can only
be arbitrarily sharp, see \citep{zetterstrom2025investigations} for more details.
The code is:

\begin{verbatim}
checksharpSVbound(whichEst = "RR_tot",
        sens = sensparSV,
        pY1 = c(0.286, 0.004))
\end{verbatim}

\begin{verbatim}
#> The lower SV bound for the risk ratio in the total population is arbitrarily sharp.
#> See vignette for details.
\end{verbatim}

\begin{verbatim}
#> The upper SV bound for the risk ratio in the total population is arbitrarily sharp.
#> See vignette for details.
\end{verbatim}

The output is a string that states whether the SV bounds are sharp or not.
For these settings, both the lower and upper SV bounds are arbitrarily sharp.

\subsection{\texorpdfstring{\texttt{AFbound()}}{AFbound()}}\label{afbound}

The AF bound is calculated using the function \texttt{AFbound()}. The first argument is
the causal estimand of interest (\texttt{"RR\_tot"}, \texttt{"RD\_tot"}, \texttt{"RR\_sub"} or
\texttt{"RD\_sub"}). The second argument is \texttt{outcome}, where the user inputs
either the numeric vector with the observed outcome variable or a vector with
the conditional outcome probabilities, \(P(Y=1|T=1,I_S=1)\) and \(P(Y=1|T=0,I_S=1)\).
The third argument is \texttt{treatment}, where the user inputs either the numeric
vector with the observed treatment variable or a vector with the conditional
treatment probabilities, \(P(T=1|I_S=1)\) and \(P(T=0|I_S=1)\). The fourth argument
is \texttt{selection} where the user inputs either the numeric vector with the observed
selection variable, or the selection probability, \(P(I_S=1)\). Its default value
is \texttt{NULL} since it is only required when the causal estimands in the total
population are of interest. If the subpopulation is of interest and
\texttt{selection\ =\ NULL}, the outcome and treatment vectors must only include the
selected subjects. The output is the lower and upper AF bounds. The code and
output are:

\begin{verbatim}
attach(zika_learner)

AFbound(whichEst = "RR_tot",
        outcome = mic_ceph[sel_ind == 1],
        treatment = zika[sel_ind == 1],
        selection = mean(sel_ind))
\end{verbatim}

\begin{verbatim}
#>      [,1]             [,2]  
#> [1,] "AF lower bound" 0     
#> [2,] "AF upper bound" 454.09
\end{verbatim}

Similar to before, \texttt{whichEst\ =\ "RR\_tot"}. Furthermore, the outcome and treatment
variables are microcephaly and zika. The selection probability is specified
since the other variables are restricted to those subjects with \(I_S=1\). The
output is the lower and upper AF bounds, which are 0 and 454.09 in the zika
example.

If the raw data is not available, one can input the conditional probabilities
instead. In this example, these probabilities are:

\begin{verbatim}
AFbound(whichEst = "RR_tot",
        outcome = c(0.286, 0.004),
        treatment = c(0.002, 0.998),
        selection = mean(sel_ind))
\end{verbatim}

\begin{verbatim}
#>      [,1]             [,2]  
#> [1,] "AF lower bound" 0     
#> [2,] "AF upper bound" 435.14
\end{verbatim}

The difference in these two examples comes from rounding errors in the input
probabilities.

\subsection{\texorpdfstring{\texttt{GAFbound()}}{GAFbound()}}\label{gafbound}

The GAF bound is calculated using the function \texttt{GAFbound()}. The first argument
is the causal estimand of interest (\texttt{"RR\_tot"}, \texttt{"RD\_tot"}, \texttt{"RR\_sub"} or
\texttt{"RD\_sub"}). The second argument is \texttt{sens}, which is either output from
\texttt{sensitivityparametersM()}, a named list, or a data frame with column names
``parameter'' and ``value'', which includes the sensitivity parameters. This input
is optional as the sensitivity parameters can also be specified directly in the
function with the below arguments. The third and fourth arguments are \texttt{M} and
\texttt{m} which are the two sensitivity parameters for the GAF bound. The sensitivity
parameters can either be calculated using \texttt{sensitivityparametersM()}, or found
elsewhere. For sensitivity parameters found elsewhere, \texttt{GAFbound()} is not
restricted to the generalized M-structure. However, the necessary assumptions
for the GAF bound must still be fulfilled \citep{zetterstrom2024}. The fifth argument
is \texttt{outcome}, where the user inputs either the observed numeric vector with the
outcome variable or a vector with the conditional outcome probabilities,
\(P(Y=1|T=1,I_S=1)\) and \(P(Y=1|T=0,I_S=1)\). The sixth argument is \texttt{treatment},
where the user inputs either the observed numeric vector with the treatment
variable or a vector with the conditional treatment probabilities,
\(P(T=1|I_S=1)\) and \(P(T=0|I_S=1)\). The seventh argument is \texttt{selection} where the
user can either input the observed selection vector or selection probability.
Its default value is \texttt{NULL} since it is only required when the causal estimands
in the total population are of interest. If the subpopulation is of interest and
\texttt{selection\ =\ NULL}, the outcome and treatment vectors must only include the
selected subjects. The output is the lower and upper GAF bounds. The code and
output are:

\begin{verbatim}
GAFbound(whichEst = "RR_tot",
         M = 0.4502,
         m = 0.002, 
         outcome = mic_ceph[sel_ind == 1],
         treatment = zika[sel_ind == 1],
         selection = mean(sel_ind))
\end{verbatim}

\begin{verbatim}
#>      [,1]              [,2]  
#> [1,] "GAF lower bound" 0.01  
#> [2,] "GAF upper bound" 147.42
\end{verbatim}

\begin{verbatim}
GAFbound(whichEst = "RR_tot",
         sens = sensparGAF, 
         outcome = mic_ceph[sel_ind == 1],
         treatment = zika[sel_ind == 1],
         selection = mean(sel_ind))
\end{verbatim}

\begin{verbatim}
#>      [,1]              [,2]  
#> [1,] "GAF lower bound" 0.01  
#> [2,] "GAF upper bound" 147.42
\end{verbatim}

Similar to before, the estimand of interest is the risk ratio in the total
population, \texttt{whichEst\ =\ "RR\_tot"}. The sensitivity parameters are the
output from \texttt{sensitivityparametersM()}. Furthermore, the outcome and treatment
variables are microcephaly and zika. The selection probability is specified
since the other variables are restricted to those subjects with \(I_S=1\). The
output is the lower and upper GAF bounds, which are 0.01 and 147.42 in the
zika example. If the raw data is not available, one can input the conditional
probabilities instead, similar to the AF bound.

\subsection{\texorpdfstring{\texttt{CAFbound()}}{CAFbound()}}\label{cafbound}

The CAF bound is calculated using the function \texttt{CAFbound()}. The first argument
is the causal estimand of interest (\texttt{"RR\_tot"}, \texttt{"RD\_tot"}, \texttt{"RR\_sub"} or
\texttt{"RD\_sub"}). The second and third arguments are \texttt{M} and \texttt{m} which are the two
sensitivity parameters for the CAF bound. The fourth argument is \texttt{outcome},
where the user inputs either the numeric vector with the observed outcome
variable or a vector with the conditional outcome probabilities,
\(P(Y=1|T=1,I_S=1)\) and \(P(Y=1|T=0,I_S=1)\). The fifth argument is \texttt{treatment},
where the user inputs either the numeric vector with the observed treatment
variable or a vector with the conditional treatment probabilities,
\(P(T=1|I_S=1)\) and \(P(T=0|I_S=1)\). The sixth argument is \texttt{selection} where the
user inputs either the numeric vector with the observed selection variable, or
the selection probability, \(P(I_S=1)\). Its default value is \texttt{NULL} since it is
only required when the causal estimands in the total population are of interest.
If the subpopulation is of interest and \texttt{selection\ =\ NULL}, the outcome and
treatment vectors must only include the selected subjects. The output is the
lower and upper CAF bounds. The code and output are:

\begin{verbatim}
CAFbound(whichEst = "RR_tot",
         M = 0.3,
         m = 0.005, 
         outcome = c(0.286, 0.004),
         treatment = c(0.002, 0.998),
         selection = mean(sel_ind))
\end{verbatim}

\begin{verbatim}
#>      [,1]              [,2] 
#> [1,] "CAF lower bound" 0.04 
#> [2,] "CAF upper bound" 67.78
\end{verbatim}

Similar to before, \texttt{whichEst\ =\ "RR\_tot"}. The sensitivity parameters are chosen
by the user. Furthermore, the outcome and treatment variables are microcephaly
and zika. The selection probability is specified since the other variables are
restricted to those subjects with \(I_S=1\). The output is the lower and upper
CAF bounds, which are 0.04 and 67.78 in the zika example. If the raw data is not
available, one can input the conditional probabilities instead, similar to the
AF and GAF bounds.

\section{Comparison to other software for sensitivity analysis}\label{comparison-to-other-software-for-sensitivity-analysis}

Sensitivity analysis is an important tool when the statistical analysis relies on
untestable assumptions, and there are many R packages that can be used to
perform sensitivity analyses for different types of bias and using
different methods. Here, we address some of them and compare them to
\CRANpkg{SelectionBias}.

The R packages \CRANpkg{tipr} \citep{mcgowan2022tipr} and \CRANpkg{sensemakr}
\citep{sensemakr2024} solely focus on bias that arises from unmeasured confounding
between a treatment and outcome variable (although \CRANpkg{sensemakr} calls
this ``omitted-variable bias''), and they are therefore not concerning the same
situation as \CRANpkg{SelectionBias}. The numerical bounds for multiple types of
biases presented in \citet{duarte2024automated} are accompanied by a Python package.
These bounds make fewer assumptions than the SV, sharp, and GAF bounds, and can
therefore be more conservative, and are not available in R. The R package with
most overlap is the \CRANpkg{EValue} package, and below follows a more detailed
comparison of them.

The R package \CRANpkg{EValue} provides sensitivity analysis methods for
unmeasured confounding, measurement error, selection bias, and combinations of
them \citep{ding2017sensitivity, mathur2020sensitivity}. However, \CRANpkg{EValue}
only provides one method of sensitivity analysis for selection bias: the SV
bound (and other versions of the SV bound that require more information and
assumptions) and the accompanying selection E-value. The selection E-value is a
summary measure that calculates how large the sensitivity parameters must be in
order to explain away an effect under the assumption that they all are equal,
i.e.~a type of Cornfield inequality \citep{cornfield1959smoking}.

\CRANpkg{SelectionBias} on the other hand includes four other bounds (in
addition to the SV bound): the sharp, GAF, CAF, and AF bounds. The user
therefore has more options for their sensitivity analyses for selection bias.
This is important since the different bounds are valid under different
assumptions, meaning that they can be used in various cases. As a result, the
package \CRANpkg{SelectionBias} can cover more cases than the \CRANpkg{EValue}
package. The sensitivity parameters required in the SV bound may in some cases
also be difficult to specify, and \CRANpkg{SelectionBias} can calculate the
sensitivity parameters for a user-specified DGP whereas \CRANpkg{EValue}
requires the user to specify them. This
DGP may be easier to specify than the sensitivity parameters in some cases.
\CRANpkg{SelectionBias} can also check sharpness of the SV bound, i.e.~check if
it is possible that the bias is as large as the bound or if it is too conservative.

To summarize, if the assumptions for the SV bound are fulfilled, and the user
is confident in specifying the sensitivity parameters for the SV bounds, or
wants to calculate bounds for other types of biases as well, the package
\CRANpkg{EValue} provides the desired functionality. On the other hand, if the
assumptions for the SV bound are not valid, or the user is more confident in
specifying a DGP, or wishes to compare different bounds for the selection bias,
\CRANpkg{SelectionBias} provides the required functionality. Including several
bounds in one framework, together with more comprehensive numerical and data
examples, allows \CRANpkg{SelectionBias} to facilitate comparisons between
bounds. The different features of these
packages are summarized in Table \ref{tab:packages}.

\newpage

\begin{longtable}[]{@{}
  >{\centering\arraybackslash}p{(\linewidth - 12\tabcolsep) * \real{0.1088}}
  >{\centering\arraybackslash}p{(\linewidth - 12\tabcolsep) * \real{0.2789}}
  >{\centering\arraybackslash}p{(\linewidth - 12\tabcolsep) * \real{0.1701}}
  >{\centering\arraybackslash}p{(\linewidth - 12\tabcolsep) * \real{0.1361}}
  >{\centering\arraybackslash}p{(\linewidth - 12\tabcolsep) * \real{0.1020}}
  >{\centering\arraybackslash}p{(\linewidth - 12\tabcolsep) * \real{0.1224}}
  >{\centering\arraybackslash}p{(\linewidth - 12\tabcolsep) * \real{0.0816}}@{}}
\caption{\label{tab:packages} Summary of the features of the different packages.}\tabularnewline
\toprule\noalign{}
\begin{minipage}[b]{\linewidth}\centering
Package
\end{minipage} & \begin{minipage}[b]{\linewidth}\centering
Calculate sensitivity parameters from DGP
\end{minipage} & \begin{minipage}[b]{\linewidth}\centering
Multiple bounds
\end{minipage} & \begin{minipage}[b]{\linewidth}\centering
Additional SV bounds
\end{minipage} & \begin{minipage}[b]{\linewidth}\centering
Summary measure
\end{minipage} & \begin{minipage}[b]{\linewidth}\centering
Evaluate sharpness
\end{minipage} & \begin{minipage}[b]{\linewidth}\centering
Other biases
\end{minipage} \\
\midrule\noalign{}
\endfirsthead
\toprule\noalign{}
\begin{minipage}[b]{\linewidth}\centering
Package
\end{minipage} & \begin{minipage}[b]{\linewidth}\centering
Calculate sensitivity parameters from DGP
\end{minipage} & \begin{minipage}[b]{\linewidth}\centering
Multiple bounds
\end{minipage} & \begin{minipage}[b]{\linewidth}\centering
Additional SV bounds
\end{minipage} & \begin{minipage}[b]{\linewidth}\centering
Summary measure
\end{minipage} & \begin{minipage}[b]{\linewidth}\centering
Evaluate sharpness
\end{minipage} & \begin{minipage}[b]{\linewidth}\centering
Other biases
\end{minipage} \\
\midrule\noalign{}
\endhead
\bottomrule\noalign{}
\endlastfoot
EValue & No & No & Yes & Yes & No & Yes \\
& & & & & & \\
& & Yes: & & & & \\
& & sharp & & & & \\
SelectionBias & Yes & GAF & No & No & Yes & No \\
& & CAF & & & & \\
& & AF & & & & \\
\end{longtable}

\section{Conclusion}\label{conclusion}

To perform sensitivity analyses for selection bias, an option is to calculate
bounds. For applied researchers, we present the R package \CRANpkg{SelectionBias}.
The \CRANpkg{SelectionBias} package provides functions for calculating sensitivity
parameters and bounds previously proposed in the literature. In detail, the
package includes a function that calculates the sensitivity parameters for the
SV, sharp, and GAF bounds for the generalized M-structure specified by the
user, functions for calculating the SV, sharp, AF, GAF, and CAF bounds, and a
function for checking whether the calculated SV bound is sharp
\citep{smith2019bounding, zetterstrom2025investigations, zetterstrom2024, zetterstrom2022selection}.
For users to
become familiar with the functions, the R package also includes a simulated
dataset, \texttt{zika\_learner}, and the DGP the dataset is simulated from. The
simulated data emulates a register-based cohort study, inspired by a case-control
study investigating the effect of zika virus on microcephaly in Brazil
\citep{de2018association}. Hopefully, \CRANpkg{SelectionBias} can be a complement to
the already existing package \CRANpkg{EValue}. \CRANpkg{EValue} is a good option
when the user has knowledge of the sensitivity parameters, or when the user wants to
address several types of biases. \CRANpkg{SelectionBias} on the other
hand can be an option if the user wants to compare several different bounds or
has knowledge about the DGP. Furthermore, with \CRANpkg{SelectionBias}, the
user can perform the calculations for the sensitivity analysis for many possible
designs and different bounds, and thereby gain a broader knowledge of the
possible magnitude of the selection bias. The package is continuously updated with
new sensitivity analyses and additional functions. For instance, bootstrap is
expected to be included to account for sampling variability.

\section*{Appendix}\label{appendix}
\addcontentsline{toc}{section}{Appendix}

\subsection*{Underlying assumptions for the SV, sharp, and GAF bounds}\label{underlying-assumptions-for-the-sv-sharp-and-gaf-bounds}
\addcontentsline{toc}{subsection}{Underlying assumptions for the SV, sharp, and GAF bounds}

The SV, sharp, and GAF bounds are valid under additional assumptions. The bounds
are derived under two different conditional independence assumptions for the
causal model, depending on whether the total or subpopulation estimands are of
interest:

\begin{enumerate}
\def\labelenumi{\arabic{enumi}.}
\item
  \emph{(Total population estimands \(\beta_R\) and \(\beta_D\))
  For some unmeasured variable(s) \textit{U}: \(Y \perp\hskip -7pt \perp I_S|(T=t,U=u)\),
  for \(t=0,1\).}
\item
  \emph{(Subpopulation estimands \(\beta_{R_S}\) and \(\beta_{D_S}\))
  For some unmeasured variable(s) \textit{U}: \(Y(t) \perp\hskip -7pt \perp T|(I_S=1,U=u)\),
  for \(t=0,1\).}
\end{enumerate}

The assumptions differ since their purpose is to provide conditions for unbiased
estimation of the estimands of interest under selection if the unmeasured
variable was observed. The difference arises from either i) a lack of
generalization of a causal effect from a subpopulation to the total population,
or ii) a violation of conditional exchangeability in the subpopulation, i.e.,
\(Y(t)\perp\hskip -7pt \perp T|X,I_S=1\) does not hold. There are however structures
in which both assumptions are fulfilled, for example the generalized M-structure
in Figure \ref{fig:genM}.

\subsection*{The bounds}\label{the-bounds}
\addcontentsline{toc}{subsection}{The bounds}

All bounds have the structure
\[
LB<\beta<UB.
\]
The SV bounds for the risk ratio in the total population are
\[
\frac{\beta_R^{obs}}{BF_{11} BF_{00}} \leq  \beta_R \leq \beta_R^{obs} \cdot BF_{10}BF_{01}.
\]
The SV bounds for the risk difference in the total population are
\[
\beta_D^{obs} - (BF_{11} - \frac{P(Y=1|T=1,I_S=1)}{BF_{11}} + P(Y=1|T=0,I_S=1)\cdot BF_{00}) \leq  \beta_D \leq
\]
\[
\beta_D^{obs} + (BF_{01} - \frac{P(Y=1|T=0,I_S=1)}{BF_{01}} + P(Y=1|T=1,I_S=1)\cdot BF_{10}).
\]
The SV bounds for the risk ratio in the subpopulation are
\[
\frac{\beta_R^{obs}}{BF_{1}} \leq  \beta_{R_S} \leq \beta_R^{obs} \cdot BF_{0}.
\]
The SV bounds for the risk difference in the subpopulation are
\[
\beta_D^{obs} - \max(P(Y=1|T=0,I_S=1)\cdot (BF_1-1),P(Y=1|T=1,I_S=1)\cdot (1-1/BF_1)) \leq \beta_{D_S} \leq 
\]
\[
\beta_D^{obs} + \max(P(Y=1|T=1,I_S=1)\cdot (BF_0-1),P(Y=1|T=0,I_S=1)\cdot (1-1/BF_0)).
\]
The alternative SV bounds for the risk difference in the subpopulation are
\[
\beta_D^{obs} - [P(Y=1|T=0,I_S=1)\cdot P(T=1|I_S=1) \cdot (BF_1-1) + P(Y=1|T=1,I_S=1)\cdot P(T=0|I_S=1) \cdot (1-1/BF_1)] 
\]
\[
\leq \beta_{D_S} \leq
\]
\[
\beta_D^{obs} + [P(Y=1|T=1,I_S=1) \cdot P(T=0|I_S=1) \cdot (BF_0-1) + P(Y=1|T=0,I_S=1)\cdot P(T=1|I_S=1) \cdot (1-1/BF_0)].
\]\\
The sharp bounds are presented as bounds for the potential outcome probabilities
which can then be used to construct bounds for the causal estimands. The bounds
for the total population are
\[
l_t=p(Y=1\,|\,T=t,I_S=1)\{p(I_S=1\,|\,T=t)+p(I_S=0\,|\,T=t)/BF_{t1}\}
\]
and
\[
u_t=p(Y=1\,|\,T=t,I_S=1)\times\left[p(I_S=1\,|\,T=t)+p(I_S=0\,|\,T=t)\times\min(BF_{t0},1/p(Y=1\,|\,T=t,I_S=1))\right],
\]
which results in the following bounds for \(p(Y(t)=1)\):
\[
l_t\leq p(Y(t)=1)\leq u_t.
\]
Upper (lower) bounds for the causal risk ratio and causal risk difference are
found by combining \(l_0\) (\(l_1\)) and \(u_1\) (\(u_0\)). The sharp bounds for the
subpopulation potential outcome probabilities are
\[
l_t'=p(Y=1\,|\,T=t,I_S=1)[p(T=t\,|\,I_S=1)+p(T=1-t\,|\,I_S=1)/BF_{t}]
\]
and
\[
u_t'=p(Y=1\,|\,T=t,I_S=1)\times\left[p(T=t\,|\,I_S=1)+p(T=1-t\,|\,I_S=1)\times\min\{BF_{(1-t)},1/p(Y=1\,|\,T=t,I_S=1)\}\right],
\]
which results in the following bounds for \(p(Y(t)=1\,|\,I_S=1)\):
\[
l_t'\leq p(Y(t)=1\,|\,I_S=1)\leq u_t'
\]
which can then be used to construct bounds for the causal risk ratio and
causal risk difference.

For the GAF, CAF, and AF bounds the potential outcome probabilities in the
total population can be bounded as

\[
P(Y=1,T=t,S=1)+\left[1-P(T=t,S=1)\right]\cdot m_T < P(Y(t)=1) <
\]
\[
P(Y=1,T=t,S=1)+\left[1-P(T=t,S=1)\right]\cdot M_T.
\]
The GAF bounds are calculated by using
\[
    m_T=\min_{t,u}P(Y=1|T=t,U=u)
\]
and
\[
    M_T= \max_{t,u}P(Y=1|T=t,U=u),
\]
the CAF bounds are calculated by using
\[
    m_T=\min_t P(Y=1|T=t,I_S=0)
\]
and
\[
    M_T=\max_t P(Y=1|T=t,I_S=0),
\]
and the AF bounds are calculated by using \(m_T=0\) and \(M_T=1\). These bounds
for the potential outcome probabilities can then be used to construct bounds
for the causal risk ratio and causal risk difference.

For the GAF, CAF, and AF bounds the potential outcome probabilities in the
subpopulation can be bounded as
\[
    P(Y=1,T=t|I_S=1) + P(T=1-t|I_S=1)\cdot m_S < P(Y(t)=1|S=1) < 
\]
\[
    P(Y=1,T=t|S=1) + P(T=1-t|I_S=1)\cdot M_S.
\]
For the GAF bounds,
\[
    m_S=\min_{t,u}P(Y=1|T=t,U=u,S=1)
\]
and
\[
    M_S =\max_{t,u}P(Y=1|T=t,U=u,S=1),
\]
for the CAF bounds
\[
    m_S=\min_t P(Y(t)=1|T=1-t,I_S=1)
\]
and
\[
    M_S=\max_t P(Y(t)=1|T=1-t,I_S=1),
\]
and for the AF bounds \(m_S=0\) and \(M_S=1\). These bounds for the potential
outcome probabilities can then be used to construct bounds for the causal risk
ratio and causal risk difference.

\bibliography{ZetterstromWaernbaum.bib}

\address{%
Stina Zetterstrom\\
Uppsala University\\%
Department of Statistics\\ Uppsala, Sweden\\
\url{https://www.statistik.uu.se/?languageId=1}\\%
\textit{ORCiD: \href{https://orcid.org/0000-0003-0730-9425}{0000-0003-0730-9425}}\\%
\href{mailto:stina.zetterstrom@gmail.com}{\nolinkurl{stina.zetterstrom@gmail.com}}%
}

\address{%
Ingeborg Waernbaum\\
Uppsala University\\%
Department of Statistics\\ Uppsala, Sweden\\
\url{https://www.statistik.uu.se/?languageId=1}\\%
\textit{ORCiD: \href{https://orcid.org/0000-0002-4457-5311}{0000-0002-4457-5311}}\\%
\href{mailto:ingeborg.waernbaum@statistics.uu.se}{\nolinkurl{ingeborg.waernbaum@statistics.uu.se}}%
}

\end{article}

\end{document}